\documentclass{article}
\usepackage{ijcai25}

\usepackage{times}
\usepackage{soul}
\usepackage{url}
\usepackage[hidelinks]{hyperref}
\usepackage[utf8]{inputenc}
\usepackage[small]{caption}
\usepackage{graphicx}
\usepackage{amsmath}
\usepackage{amsthm}
\usepackage{booktabs}
\usepackage{algorithm}
\usepackage{algorithmic}
\usepackage{multirow}
\usepackage[switch]{lineno}

\title{Evaluation and Verification of Physics-Informed Neural Models of the Grad-Shafranov Equation}%% with Formal Verification}

\author{
Fauzan Nazranda Rizqan
\and
Matthew Hole\and
Charles Gretton
\affiliations
Australian National University\\
\emails
\{u7510136, matthew.hole, charles.gretton\}@anu.edu.au
}

\begin{document}

\maketitle

\begin{abstract}
Our contributions are motivated by fusion reactors that rely on maintaining magnetohydrodynamic (MHD) equilibrium, where the balance between plasma pressure and confining magnetic fields is required for stable operation. In axisymmetric tokamak reactors in particular, and under the assumption of toroidal symmetry, this equilibrium can be mathematically modelled using the Grad-Shafranov Equation (GSE). Recent works have demonstrated the potential of using Physics-Informed Neural Networks (PINNs) to model the GSE. Existing studies did not examine realistic scenarios in which a single network generalizes to a variety of boundary conditions. Addressing that limitation, we evaluate a PINN architecture that incorporates boundary points as network inputs. Additionally, we compare PINN model accuracy and inference speeds with a Fourier Neural Operator (FNO) model. Finding the PINN model to be the most performant, and accurate in our setting, we use the network verification tool Marabou to perform a range of verification tasks. 
Although we find some discrepancies between evaluations of the networks natively in PyTorch, compared to via Marabou, we are able to demonstrate useful and practical verification workflows. Our study is the first investigation of verification of such networks.
%%
%%We validate physical constraints by searching through a range of flux values, with only minor discrepancies caused by floating-point errors. This novel application of formal verification highlights its potential for improving reliability in fusion research, enabling real-time control and robust surrogate modelling. 
\end{abstract}

\section{Introduction}
Fusion energy holds immense potential due to its capability to provide a nearly inexhaustible, safe, and environmentally friendly power source \cite{Smith_Cowley_2010_ref1}.
It offers a promising solution to the world’s growing energy demands while reducing our reliance on fossil fuels and minimizing environmental impact. Harnessing this potential requires designing fusion reactors that create and maintain the state of magnetohydrodynamics (MHD) equlibrium. \cite{ref2}. The state of MHD equilibrium is achievable through careful manipulation of the magnetic fields in order that they balance the large pressure gradient force inside reactors. For axisymmetric tokamaks, the equilibrium can be mathematically modelled through the \textit{Grad-Shafranov Equation} (GSE) \cite{grad_rubin_1958_ref3}. While past research have demonstrated the possibilities of solving the GSE using various numerical methods, iterative solvers can get computationally expensive, especially resolving highly shaped geometry.  This poses a challenge for applications requiring real-time control diagnostics. Several studies have demonstrated the possibility of leveraging machine learning (ML) to solve the GSE, both as a complement to EFIT \cite{lao_kruger_c_ref4} and independently. However, these approaches heavily rely on extensive datasets for accurate modeling, which can be a limiting factor, particularly when experimental data is scarce or incomplete.

Versatile and adaptive models are increasingly demanded for enhancing real-time control and achieving accurate plasma reconstruction while maintaining efficiency. This paper seeks to address this challenge by extending the capabilities of artificial neural models of GSE, and in particular PINN models, to improve their generality and thereby, applicability. We also evaluate Fourier Neural Operators, as an alternative to PINNs in this setting. Finally, we explore the feasibility of verification workflows using the contemporary network verification tool Marabou. Our main contributions are summarized as follows: 

\begin{enumerate}
    \item \textbf{Evaluation of Physics-Informed Neural Models for Arbitrary Fixed Boundaries.} We extended the original PINNs approach for the GSE \cite{jang_kaptanoglu_gaur_pan_landreman_dorland_2023_ref6} to accept arbitrary 2D boundaries as inputs and evaluated its generalizability in reconstructing equilibrium in real-time. Additionally, we implemented Fourier Neural Operators (FNOs) as an alternative method for solving the GSE. Both approaches were rigorously compared in terms of accuracy in predicting magnetic flux surfaces, training time, and inference speed.
    
    \item \textbf{Integrated Model Checking Analysis for PINNs.} We perform a range of network verification tasks using the \textit{Marabou}~\cite{wu2024marabou20versatileformal_ref9}, a general purpose model-checking tool for evaluating the properties of artificial neural networks. We are able to verify and validate the robustness of a PINN against meaningful perturbations. Using Marabou, we are able to identify and address potential flaws and errors in a learned model. Ultimately, we seek not only to advance our understanding of the network's robustness, but also to explore the feasibility of applying model checking to PINNs.
\end{enumerate}

\section{Related Work}
\subsection{Deep Learning for Plasma Equilibrium}

The GSE is historically solved using a variety of numerical methods executed on CPUs~\cite{orozco_dougar-zhabon_2019_ref10,ref11,howell_c.r_ref12,ref13}.
In recent times researchers have been investigating the use of GPUs to achieve relatively fast evaluations, particularly in the context of artificial neural networks.
Network-based solutions have been developed that fit such equations using data from instrument telemetry~\cite{joung_kim_kwak_bak_lee_han_kim_lee_kwon_ghim_2019_ref15,ref16}, with separate networks trained to fit and measure physicality of results. 
PINNs are also studied and motivated in our setting, and are a compelling solution concept because they explicitly capture the physical laws. 
\citeauthor{jang_kaptanoglu_gaur_pan_landreman_dorland_2023_ref6}~(\citeyear{jang_kaptanoglu_gaur_pan_landreman_dorland_2023_ref6}) develop a PINN for GSE with fixed-boundary conditions, directly employing the GSE as a governing loss.
An appealing aspect of this approach is the ability to train on synthetic data, and the speed of network evaluation which is amenable to real-time applications, such as control.

Neural Operators~\cite{ref17} provide an alternative and adaptable architecture for modelling physical systems, with Fourier Neural Operators (FNOs) in particular offering an attractive computationally efficient approach here.
\citeauthor{li_nikolakovachki_kamyarazizzadenesheli_liu_bhattacharya_stuart_animaanandkumar_2020_ref7}~(\citeyear{li_nikolakovachki_kamyarazizzadenesheli_liu_bhattacharya_stuart_animaanandkumar_2020_ref7}) argue that FNOs offer a speed and computational cost advantage over PINNs in circumstances where evaluation is over a broad class of parameters. We explore both FNO and PINN models for modelling GSE.   

\subsection{Verification of Neural Network}

Tokamak fusion power plants will be susceptible to high-cost failure modes and safety criticality, and this motivates rigorous formal verification of their control systems. If artificial neural models are to be developed for use in tokamak control, network verification is a hurdle that must be overcome in order to guarantee robustness, safety, and correctness~\cite{ref21}. Treating deep learning generally, several studies have shown that despite being effective, network models can be vulnerable to adversarial ``attacks'' corresponding to synthetic inputs that expose important network model flaws~\cite{ref18,ref19,ref20}. Several model checking systems have been developed for network verification, including Verinet~\cite{ref22,ref23},  $\alpha$-$\beta$-CROWN \cite{ref24,ref25,ref26,ref27,ref28,ref29}, Marabou \cite{katz_huang_ibeling_julian_lazarus_lim_shah_thakoor_wu_zelji_etal._2019_ref8}, and Neurify \cite{ref30}.
Marabou, which began as a reimplementation of Reluplex~\cite{ref31}, is a simple highly flexible verification tool that recently integrated support, via its CEGAR subsolver, for verifying network with non-linear activation functions~\cite{katz_huang_ibeling_julian_lazarus_lim_shah_thakoor_wu_zelji_etal._2019_ref8,wu2024marabou20versatileformal_ref9}.
We have evaluated Marabou in solving a range of PINN verification tasks, noting that the PINN models of GSE feature $\tanh$ activations.

% Print manuscripts two columns to a page, in the manner in which these
% instructions are printed. The exact dimensions for pages are:
% \begin{itemize}
%     \item left and right margins: .75$''$
%     \item column width: 3.375$''$
%     \item gap between columns: .25$''$
%     \item top margin---first page: 1.375$''$
%     \item top margin---other pages: .75$''$
%     \item bottom margin: 1.25$''$
%     \item column height---first page: 6.625$''$
%     \item column height---other pages: 9$''$
% \end{itemize}

% All measurements assume an 8-1/2$''$ $\times$ 11$''$ page size. For
% A4-size paper, use the given top and left margins, column width,
% height, and gap, and modify the bottom and right margins as necessary.

\section{Physics-Informed Neural Models}
\subsection{Dataset Generation}

Our work aims to evaluate the performance of the models under controlled conditions. While using actual experimental data is possible, for simplicity and consistency here we use an analytical model to generate  training and testing data. We proceed with plasma boundary shapes that are given according to the parametric equation used by~\citeauthor{ref32_cerfon_freidberg}~(\citeyear{ref32_cerfon_freidberg}). 
The equation describes a smooth elongated “D”-shaped cross-section, taking the following form:
\begin{equation}
\label{eq:boundary}
R = 1 + \epsilon \cos\left(\tau + \arcsin(\delta) \sin(\tau)\right),
\quad
Z = \epsilon \kappa \sin(\tau)
\end{equation}
Here, \(\epsilon\) is the inverse aspect ratio, \(\kappa\) is the elongation, \(\delta\) is the triangularity, and \(\tau \in [0, 2\pi]\). Each set of boundary is described to the network as an input, comprising 100 points spaced uniformly along the contour of the boundary shape.

For our experiments, we trained our models over a sampled range of boundary parameters. Specifically, the parameters $\epsilon$, $\kappa$, and $\delta$ were sampled uniformly within the ranges 0.30–0.95, 1.5–3.0, and -0.3–0.5, respectively, with 10 samples per parameter. Additionally, 6 sets of boundaries with a null divertor and 5 iterations of the \(P\) parameter, varied from 0.0 to 1.0, were included. 

\begin{figure}
    \centering
    \includegraphics[width=0.75\linewidth]{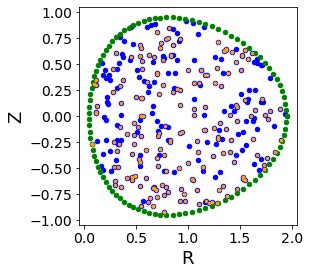}
    \caption{Example of adaptive resampling visualization: Green dots represent boundary points, blue dots indicate randomly generated points, and orange dots highlight points with higher errors that will be prioritized for the next iteration.}
    \label{fig:sampling}
\end{figure}

Our network enforces physical properties through loss functions by feeding points into the governing physics equations and penalizing any output deviations from the expected physical behavior. Thus, we required points within the boundary, referred to as {\em grid points}, to enable the network to learn the underlying physics of the MHD equilibrium state. For this, we implemented an adaptive resampling strategy, using a \textit{Probability Density Function} (PDF) as proposed in~\cite{ref33_adaptive}. In training, grid points are sampled in proportion to the magnitude of the incumbent network error at those points. An illustration is shown in Figure \ref{fig:sampling}, where green dots represent the boundary points, blue dots indicate randomly generated points, and orange dots highlight the resampled points with the higher error.

\begin{figure*} 
    \centering
    \includegraphics[width=0.85\linewidth]{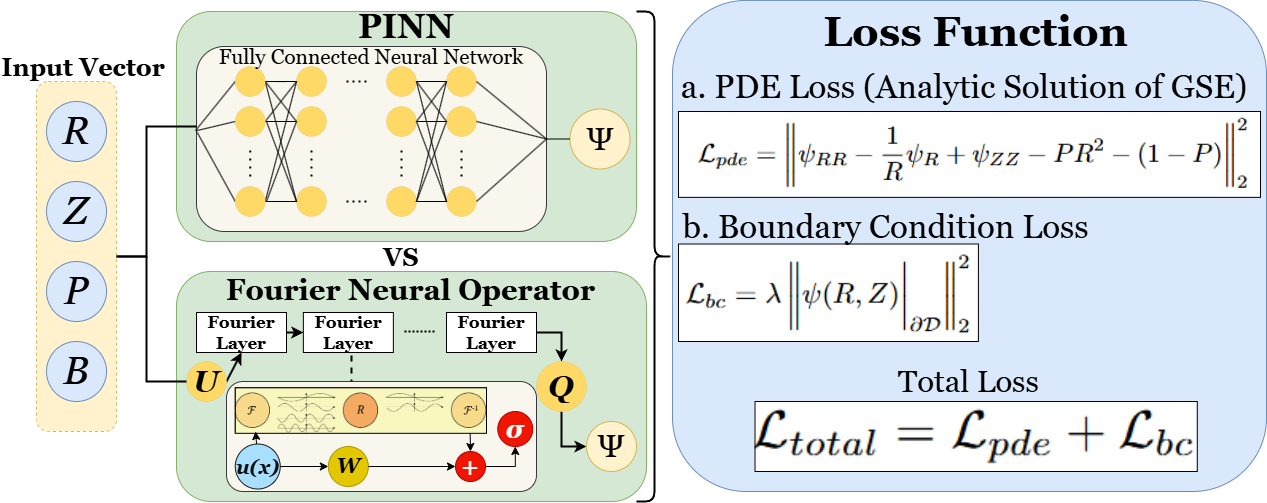}
    \caption{Diagram illustrating the network used for evaluation, showcasing the PINN and FNO architectures.}
    \label{fig:network_architecture}
\end{figure*}

\subsection{Training Regime}
The GSE are typically solved using numerical methods. However, as this work focuses on developing a network suitable for evaluation, our training loss regime follows the approach outlined in \cite{jang_kaptanoglu_gaur_pan_landreman_dorland_2023_ref6}, where the models are exposed only to analytical solutions of the GSE as provided in \cite{ref32_cerfon_freidberg}.

Generally, The GSE can be expressed as a poloidal flux function $\psi$($r$,$z$) such that: 
\begin{equation}
\psi_{rr} - \frac{1}{r} \psi_{r} + \psi_{zz} + \mu_0 r^2 \frac{dp(\psi)}{d\psi} + \frac{1}{2} \frac{dF^2(\psi)}{d\psi} = 0
\end{equation}
This general form includes nonlinearity in the pressure profile \(p(\psi)\) and the poloidal current profile \(F(\psi)\), which require numerical solutions. However, following the approach prescribed in \cite{ref32_cerfon_freidberg} and using a certain Solov'ev assumption, we assume linearity for \(p(\psi)\) and \(F^2(\psi)\). Furthermore, by applying a rescaling defined as \(R = \frac{r}{R_0}\), \(Z = \frac{z}{R_0}\), \(\Psi = \frac{\psi}{\psi_0}\), and \(\psi_0 = R_0^2(A + CR_0^2)\) for constant \(A\) and \(C\), the expression can be simplified, leading to the final governing equation:
\begin{equation}
\label{eq:pde_loss}
\Psi_{RR} - \frac{1}{R} \Psi_R + \Psi_{ZZ} - PR^2 - (1 - P) = 0
\end{equation}
Here, the additional parameter \( P \) represents the strength of the linear pressure profile and simplifies the equation by reducing the number of separate terms and parameters.

Our model enforces the physical properties of the system by incorporating the governing equation and boundary conditions as loss terms. Specifically, the PDE loss, as given in Equation \ref{eq:pde_loss}, ensures that the solution satisfies the GSE, while the boundary condition loss, ensures compliance with the prescribed fixed boundary condition. Thus, our loss functions are outlined below:
\begin{equation}
\mathcal{L}_{pde} = \left\| \psi_{RR} - \frac{1}{R} \psi_{R} + \psi_{ZZ} - PR^2 - (1 - P) \right\|_2^2 
\end{equation}
\begin{equation}
\mathcal{L}_{bc} = \lambda \left\| \psi(R, Z) \bigg|_{\partial \mathcal{D}} \right\|_2^2
\end{equation}
Here, $\partial \mathcal{D}$ is the boundary domain and $\lambda$ is a hyperparameter used to balance the contributions of the PDE residual and boundary condition errors, which we set to 100. Thus, our total loss is defined as \(\mathcal{L}_{total} = \mathcal{L}_{pde} + \mathcal{L}_{bc}\). 

\subsection{Proposed Models}
Our objective is to evaluate the network's ability  to generalize to any kinds of boundary outside of its parametric equation. Let us define a new model that directly takes the boundary points as explicit inputs to the neural network alongside the grid points. For this, we define the boundary set as:
\[
\mathcal{B} = \{(R_{bound_i}, Z_{bound_i})\}_{i=1}^{N},
\]
where \( N \) is the number of boundary points, and \( (R_{bound_i}, Z_{bound_i}) \) are the coordinates of these points. And so, we can define our model \(\mathcal{N}\), parameterized by its learnable parameters $\theta$, as:

\begin{equation}
\label{eq:proposed_model}
\
\mathcal{N}(R, Z, P, \mathcal{B}; \theta) = \psi(R, Z, P, \mathcal{B}; \theta),
\
\end{equation}

This formulation allows the network to learn the relationship between the grid points and the boundary. This approach enables a single trained model to produce solutions for arbitrary boundaries that are not constrained by predefined parametric equations, such as those representing complex configurations like a null divertor.

Figure \ref{fig:network_architecture} illustrates the architecture of the proposed network. PINNs comprises a fully connected neural networks that incorporates the boundary points and grid points as inputs to predict the flux values $\Psi$ over the domain. The architecture is designed to enforce the governing equations and boundary conditions through the loss functions which can be expressed as a minimization problem
like so:
\begin{equation}
\begin{split}
\arg \min_{W, b} & \left\| \Psi_{RR} - \frac{1}{R} \Psi_R + \Psi_{ZZ} - PR^2 - (1 - P) \right\|_2^2 \\
& + \lambda \left\| \Psi(R, Z)\big|_{\partial \mathcal{D}} \right\|_2^2
\end{split}
\end{equation}

In this work, we adopted the same physics-informed approach for the FNO as for the PINN, enforcing the governing physics through the loss function.
The representation of the function is iteratively updated according to the following formula:
\begin{equation}
\
u_{t+1}(x) = \sigma \left( W u_t(x) + (K(\phi)u_t)(x) \right), \quad \forall x \in \mathcal{D}
\
\end{equation}
where \(u_t(x)\) is the function at iteration \(t\), \(\sigma\) represents a non-linear activation function, and \(W\) is a linear transformation. The term \(K(\phi)u_t)(x)\) is a kernel convolution operator in the Fourier space parameterized by $\phi$, which leverages the Fourier transform to capture interactions across the input space by transforming the computation into the frequency domain, such that:
\begin{equation}
\
(K(\phi)u_t)(x) = \mathcal{F}^{-1} \left( R_{\phi} \cdot \mathcal{F}(u_t) \right)(x), \quad \forall x \in \mathcal{D}
\
\end{equation}
Here, \(R_{\phi}\) is the Fourier transform of the periodic kernel function \(\kappa\). Lastly, the most updated representation is then projected back to the target dimension by a neural network $Q$, which then produces the predicted flux surfaces. 

\subsection{Training Implementation}
We maintained consistent training settings as outlined in \cite{jang_kaptanoglu_gaur_pan_landreman_dorland_2023_ref6}. 
The GSE requires second-order differentiation, thus ReLU activation is not feasible. Instead, we used \(\tanh\), which provides smoother gradients and is compatible with Marabou for verification. Furthermore, we are training a PINN that consists of three hidden layers with 20 nodes each. As for the FNO, we set the number of Fourier modes to five, reflecting the low-frequency nature of the GSE solutions. Both models are built using the PyTorch framework. For optimization, we used the Adam optimizer with a learning rate of  \( 1 \times 10^{-3} \) for the first 100 epochs to manage high-gradient variance and switched to L-BFGS-B for improved convergence in the later stages. We also chose to use 85 initial grid points, which is reduced by a factor of 0.96 every 200 epochs. 

\paragraph{Computational Resources.} The models were trained on an NVIDIA RTX 2080Ti GPU. Testing, evaluation, and model checking were performed on a laptop with an AMD Ryzen 9 6900HX processor, 16GB RAM, an NVIDIA RTX 3070Ti GPU, and WSL2 on Windows 11.

\section{Model Evaluation}
We are comparing both of our models based on their accuracy and inference speed on a given dataset. Before assessing the generalization of the models on an unseen dataset, we first confirmed their performance on the training dataset to determine their ability to capture the underlying physics and generate equilibrium states.
We find that a network can be trained to reconstruct equilibrium for multiple boundary configurations with low errors, including those with an X-point divertor.

Here, we report on the models' performance on unseen data. For this, we divide our evaluation into two tests: the first evaluates unseen data extrapolated outside the trained regions, and the second evaluates unseen data within the trained regions. To assess how well the models generalize to each boundary parameter, we created test sets where one boundary parameter is extrapolated outside the trained region while the others remain within the trained ranges. For accuracy assessment, we compared the result generated by our models with the analytical solution as our ground truth.

\subsection{Test 1: Extrapolation Beyond Trained Region}
For this test, we prepared 100 sets of boundary configurations for extrapolated $\epsilon$ varied from 0.95 to 1.00,  $\kappa$ varied from 3.01 to 5.00, and 50 sets of extrapolated $\delta$ is varied under two conditions: from 0.6 to 1.00 for positive values and from -0.6 to -0.9 for negative values. We assess the errors based on the analytic solutions and the average inference time for the entire dataset.

\begin{table}
    \centering
    \begin{tabular}{llrrr}
        \toprule
        \multirow{3}{*}{\textbf{Model}} & & \multicolumn{2}{c}{\textbf{Error (\%)}} & \multirow{3}{*}{\textbf{Mean Time (ms)} } \\
        \cmidrule(rr){3-4}
                      & & \textbf{Mean} & \textbf{Max} & \\
        \midrule
        \multirow{3}{*}{\textbf{PINN}} & $\epsilon$ & 3.3 & 10.7 & 1.19 \\
                                       & $\kappa$   & 2.7 & 15.2 & 1.18 \\
                                       & $\delta$   & 8.4 & 19.0 & 1.73 \\
        \midrule
        \multirow{3}{*}{\textbf{FNO}}  & $\epsilon$ & 3.4 & 10.8 & 6.21 \\
                                       & $\kappa$   & 2.6 & 13.7  & 7.09 \\
                                       & $\delta$   & 8.1 & 19.3  & 4.31 \\
        \bottomrule
    \end{tabular}
    \caption{Comparison of errors and average inference time across the entire dataset for each extrapolated boundary parameter outside the training region for PINN and FNO.}
    \label{tab:extrapolated_comparison}
\end{table}

From Table \ref{tab:extrapolated_comparison}, we can see that both models perform similarly in terms of accuracy, with average errors within a single-digit percentage. However, there are specific areas where errors grow significantly, reaching a maximum of 19.3$\%$. The models struggle the most when generalizing to the extrapolated $\delta$ parameter. Additionally, in terms of inference time, the PINN is consistently faster than the FNO by a factor of 2.5 to 5.2, which is notable given the increased complexity of the FNO architecture.

\subsection{Test 2: Generalization Within Trained Region}
Similarly, we generated the same number of test sets for each parameter as in the previous test. However, the main difference is that these test sets consist of unseen data that remain within the trained region. In Table \ref{tab:in_comparison}, we observe that compared to Test 1 the errors are relatively low in this region. The most noticeable improvement is in the maximum error, where almost all values are within a single-digit percentage, indicating better generalization of the models in this region.

\begin{table}
    \centering
    \begin{tabular}{llrrr}
        \toprule
        \multirow{3}{*}{\textbf{Model}} & & \multicolumn{2}{c}{\textbf{Error (\%)}} & \multirow{3}{*}{\textbf{Mean Time (ms)} } \\
        \cmidrule(rr){3-4}
                      & & \textbf{Mean} & \textbf{Max} & \\
        \midrule
        \multirow{3}{*}{\textbf{PINN}} & $\epsilon$ & 2.6 & 9.8 & 2.39 \\
                                       & $\kappa$   & 1.9 & 8.9 & 1.29 \\
                                       & $\delta$   & 1.7 & 7.2 & 2.94 \\
        \midrule
        \multirow{3}{*}{\textbf{FNO}}  & $\epsilon$ & 1.9 & 10.1 & 6.44 \\
                                       & $\kappa$   & 1.9 & 8.5  & 7.16 \\
                                       & $\delta$   & 1.8 & 8.6  & 4.46 \\
        \bottomrule
    \end{tabular}
    \caption{Comparison of errors and average inference time across the entire dataset for unseen data within the training region for PINN and FNO.}
    \label{tab:in_comparison}
\end{table}

\section{Formal Verification for PINNs}
\label{5.3}
\subsection{Model Preparation}
We will be using Marabou, a widely-used tool for verifying neural networks, by setting up bounded constraints across the input space domain. For this evaluation we will be using a single trained PINN model that we built using PyTorch, and  verifying the trained network's behavior within a defined set of constraints. 

\subsubsection{Parametric-PINN}
We are using the Parametric-PINN architecture that was proposed in \cite{jang_kaptanoglu_gaur_pan_landreman_dorland_2023_ref6} that takes the grid and the boundary parameters defined above as the input such that our network $\mathcal{N}$ is expressed as:
\begin{equation}
\mathcal{N}(R, Z, \epsilon, \kappa, \delta, P; \theta) = \psi(R, Z,\epsilon, \kappa, \delta, P; \theta)
\end{equation}

For this model, we have generated a training dataset that includes boundary configurations with $\epsilon$ values ranging from 0.25 to 0.75, $\kappa$ values from 1.0 to 3.0, and $\delta$ values from -0.5 to 0.5. The $P$ value will remain fixed at 0.0 to simplify the verification process and focus on the geometric parameters. More importantly, we used the $\tanh$ activation function as it is the one that is compatible with both Marabou and the GSE.
\subsubsection{PyTorch to Marabou}
\begin{figure}
    \centering
    \includegraphics[width=1.0\linewidth]{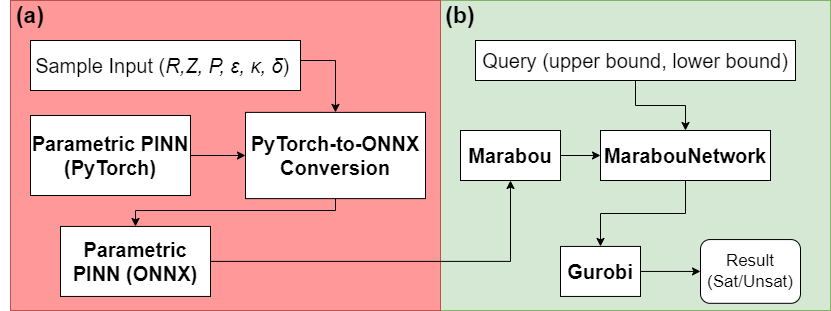}
    \caption{An overview of the process flow for model verification using Marabou. \textbf{(a)} illustrates the process of converting the model from PyTorch to ONNX, where the shape of the input sample is required. \textbf{(b)} shows the verification flow, where Marabou, given the ONNX model and queries, produces the result through Gurobi.}
    \label{fig:marabou_flow}
\end{figure}
Marabou does not directly support PyTorch models, requiring us to first convert the model into the \textit{Open Neural Network Exchange} (ONNX) format \cite{ref34}. Now, this conversion introduces slight discrepancies when evaluated in the ONNX format, on the order of \(10^{-8}\) for 32-bit precision and \(10^{-16}\) for 64-bit precision. Moreover, additional discrepancies of the same order are introduced when the model is evaluated as a Marabou object. To enable Marabou to work with our network, we are using a version of Marabou that supports the \textit{Gurobi} optimizer \cite{ref35}, which can replace the default LP solver that comes with Marabou, and that is required to  provide the solving modes needed to verify our PINNs. In this mode the solver is limited to 32-bit precision, meaning that we should account for potential discrepancies during the verification process to ensure accurate interpretation of the results. The entire flow of our model preparation is depicted in Figure \ref{fig:marabou_flow}.

\subsection{Queries Setup}

Marabou functions as a bounded constraint solver, where each verification query involves setting specific bounds on the model's inputs or outputs. For this, we set up a series of queries based on a physical properties to assess our model's adherence to the governing physics. The queries for our model architecture take the following form:
\begin{align*}
R_{\text{min}} &\leq R \leq R_{\text{max}} \\
Z_{\text{min}} &\leq Z \leq Z_{\text{max}} \\
P_{\text{min}} &\leq P \leq P_{\text{max}} \\
\epsilon_{\text{min}} &\leq \epsilon \leq \epsilon_{\text{max}} \\
\kappa_{\text{min}} &\leq \kappa \leq \kappa_{\text{max}} \\
\delta_{\text{min}} &\leq \delta \leq \delta_{\text{max}} \\
\psi_{\text{min}} &\leq \psi \leq \psi_{\text{max}}
\end{align*}
Each parameter is constrained by its lower and upper bound value according to the requirements of each query. These constraints will guide the verification process, as further elaborated in the sections below.
 
\subsection{Result}
Before assessing the tool's capability to evaluate physical constraints, we first tested the reliability of the approach to validate values through a series of simple queries. Marabou demonstrated its ability to reconstruct the inputs and outputs of our PINN model; however, as indicated above, some numerical evaluation discrepancies are observed. Thus, we relaxed the bounds for $\psi$ output by $\pm 0.001$ to balance capturing potential floating-point errors while avoiding the inclusion of irrelevant points. Here, we are focused on verifying the magnetic axis of the plasma system.

\subsubsection{Domain Exploration Test}
In this test, we are assessing Marabou's ability to explore the entire domain of a boundary. For this, we provide the expected magnetic flux value $\psi_{\text{m}}$ and task Marabou to find the point that yields the value. We conduct the search within \(R \in [0.60, 1.40]\) and \(Z \in [-0.6, 0.6]\) in a single boundary. Thus, the query will take the following form:
\begin{align*}
-0.6 &\leq Z \leq 0.6, \\
P &= 0.0, \\
\epsilon &= 0.32, \\
\kappa &= 1.70, \\
\delta &= 0.33, \\
\psi_{\text{m}} - 0.001 &\leq \psi \leq \psi_{\text{m}} + 0.001.
\end{align*}
For \(R\), we compared the performance of using multiple smaller queries against a single large query. In Table \ref{tab:multiple_results}, we can see that both approaches return different points that correspond to the same $\psi$ value. Ideally, the magnetic axis corresponds to a single point. The results obtained here are most likely due to floating-point errors. Furthermore, we observed that splitting the queries into smaller sub-domains appears to be a more efficient approach, as it reduces computational strain on the tool --  i.e., 17,642s runtime for search without subdomains, and 2,678s overall for the same query in subdomains. 

\begin{table}
    \centering
    \begin{tabular}{|l|c|r|}
        \hline
        \textbf{R Ranges} & \textbf{Result ($R$, $Z$, $\psi$)} & \textbf{Time (s)} \\
        \hline
        [0.60, 0.76] & Unsat & 0.06 \\
        \hline
        [0.76, 0.92] & Unsat & 3.15 \\
        \hline
        [0.92, 0.96] & Unsat & 0.11 \\
        \hline
        [0.96, 1.00] & Unsat & 8.93 \\
        \hline
        [1.00, 1.04] & (1.000662, -0.008196, -0.0380) & 2,663.40 \\
        \hline
        [1.04, 1.08] & Unsat & 0.12 \\
        \hline
        [1.08, 1.24] & Unsat & 3.10 \\
        \hline
        [1.24, 1.40] & Unsat & 0.06 \\
        \hline
        [0.60, 1.40] & (1.000410, -0.007501, -0.0380) & 17,642.16 \\
        \hline
    \end{tabular}
    \caption{Results of all queries across different $R$ bounds to find the magnetic flux position.}
    \label{tab:multiple_results}
\end{table}

\subsubsection{Magnetic Axis Consistency Test}
The magnetic axis is a crucial point that represents the maximum magnetic confinement, typically at the center of the innermost flux surface. In this test, we are evaluating whether the PINN model consistently predicts the magnetic axis position when varying the boundary parameter $\kappa$, ranging from 1.0 to 3.0, while keeping other parameters such as $\epsilon$ and $\delta$ fixed. The goal is to identify any potential shifts in the magnetic axis coordinates as the boundary elongates, evaluating the stability of the predicted position and its alignment with the analytical solution.

As the flux value of the magnetic axis differs with varying $\kappa$ values, to conduct this test we need to assume a linear relationship between the $\kappa$ value and the corresponding magnetic axis position. To achieve this, we will first conduct a series of data-gathering experiments using our PINN model. First, we sampled 100 values between some given range of $\kappa$ uniformly, with each given range differing by 0.10 (e.g., from 1.00 to 1.10, then 1.10 to 1.20, and so on), while keeping the other boundary parameters fixed. For each sampled $\kappa$ value, we collected the corresponding magnetic axis flux values predicted by the PINN model. With this data, we then trained a linear regression model for each range to learn the relationship between $\kappa$ and the corresponding magnetic axis coordinates by employing the following line equation model:
\begin{equation}
\
y = m\kappa + b
\
\end{equation}
Here, $y$ represents the magnetic axis flux value, $\kappa$ is the kappa parameter, $m$ is the slope indicating the rate of change in the magnetic axis as a function of $\kappa$, and $b$ is the intercept. By doing so, we can evaluate how the magnetic axis behaves across different ranges of $\kappa$, testing whether the coordinates follow the expected linear trend or exhibit any unexpected deviations. By limiting the ranges to increments of 0.10, means that we needed to run the query in Marabou 30 times to cover the full range of $\kappa$ values, from 1.00 to 3.00. However, this approach allows for minimal error in our assumption of linearity, ensuring that the magnetic axis predictions closely follow the expected trend within each smaller range.
% \begin{figure*}
%     \centering
%     \includegraphics[width=0.80\linewidth]{scaling_factor.pdf}
%     \caption{Example of the linear line relationship between the magnetic axis flux value and the $\kappa$ values for the range of 1.00 to 1.10}
%     \label{fig:linear_plot}
% \end{figure*}

We will be testing our model with this linear equation for a different set of queries. The main purpose of these queries is to determine whether it is possible to find a flux value that is higher than or equal to the magnetic axis flux, but located at a position different from the expected magnetic axis. If such a point is found, it would indicate an inconsistency in the model's prediction of the magnetic axis, suggesting that the PINN might be incorrectly predicting the location of the highest magnetic confinement. 

%%Similarly to the previous test, 
We employed an expected magnetic axis region, instead of relying on a precise point, to compensate for potential minor numerical errors and account for the variation in the magnetic axis position across different $\kappa$ values. Specifically, we will define the expected magnetic axis region as $0.99 \leq R \leq 1.01$ and $-0.02 \leq Z \leq 0.02$. Moreover, we will test it by searching through each zone around the expected regions of the magnetic axis as outlined in Table \ref{tab:zones_search}. For the queries, as mentioned above, we will be running them across $\kappa$ values ranging from 1.00 to 3.00 with increments of 0.10. With the 4 defined zones, this means we will be conducting a total of 120 queries. Therefore, the general queries will be as follows:
\begin{align*}
P &= 0.0, \\
\epsilon &= 0.32, \\
\kappa_{\text{min}} &\leq \kappa \leq \kappa_{\text{max}}, \\
\delta &= 0.33, \\
\psi &\leq m\kappa + b.
\end{align*}

\begin{table}
    \centering
    \begin{tabular}{|c|c|c|}
        \hline
        \textbf{Zone} & \textbf{R bound} & \textbf{Z bound} \\
        \hline
        Zone 1 (Left) & $R \leq 0.99$ & $Z \in [-1, 1]$ \\
        \hline
        Zone 2 (Right) & $R \geq 1.01$ & $Z \in [-1, 1]$ \\
        \hline
        Zone 3 (Below) & $R \in [0.99, 1.01]$ & $Z \leq -0.05$ \\
        \hline
        Zone 4 (Above) & $R \in [0.99, 1.01]$ & $Z \geq 0.02$ \\
        \hline
    \end{tabular}
    \caption{The divided zones used for the search queries around the magnetic axis, highlighting the distinct regions for evaluation.}
    \label{tab:zones_search}
\end{table}

\begin{table}
    \centering
    \begin{tabular}{ccr}
        \toprule
        {$\kappa$ Range} & {Counterexample Found?} & {Runtime (s)} \\
        \midrule
        1.00 - 1.40 & No & 36.21 \\
        1.40 - 1.80 & No & 33.61 \\
        1.80 - 2.20 & No & 21.07 \\
        2.20 - 2.60 & No & 13.35\\
        2.60 - 3.00 & No & 16.12\\
        \bottomrule
    \end{tabular}
    \caption{Summary of counterexamples found in each zone across different $\kappa$ ranges.}
    \label{tab:zones_results}
\end{table}

We present the summary of the results for the test in Table \ref{tab:zones_results}, where we evaluated whether any counterexamples were found in each zone across the different $\kappa$ values. As shown in the table, no counterexamples were detected, indicating that the model's predicted magnetic axis remains consistent within the defined boundaries for all tested $\kappa$ values. The query times were relatively fast, which is consistent when using Marabou to find values that are not supposed to be available in the space. Thus, we have proven that within the trained region of our $\kappa$ parameters, the coordinates of the magnetic axis are consistent. While this consistency is maintained to a certain degree---due to the relaxed constraints on the magnetic axis coordinates during verification---it ensures that the predicted axis does not deviate significantly from the expected values. This small relaxation allows for minor floating-point errors, yet confirms that the model adheres closely to the expected physical properties without exhibiting any major inaccuracies or unexpected behavior.

\begin{table}
    \centering
    \begin{tabular}{ccr}
        \toprule
        {$\kappa$ Range} & {Counterexample Found?} & {$L_2$ Error} \\
        \midrule
        4.00 - 4.10 & Yes & 0.065 \\
        4.10 - 4.20 & Yes & 0.065 \\
        4.20 - 4.30 & Yes & 0.068\\
        4.30 - 4.40 & Yes & 0.072\\
        4.40 - 4.50 & Yes & 0.071 \\
        \bottomrule
    \end{tabular}
    \caption{Summary of counterexamples found outside the trained $\kappa$ regions.}
    \label{tab:outside_zoneres}
\end{table}

\begin{figure} 
    \centering
    \includegraphics[width=1.00\linewidth]{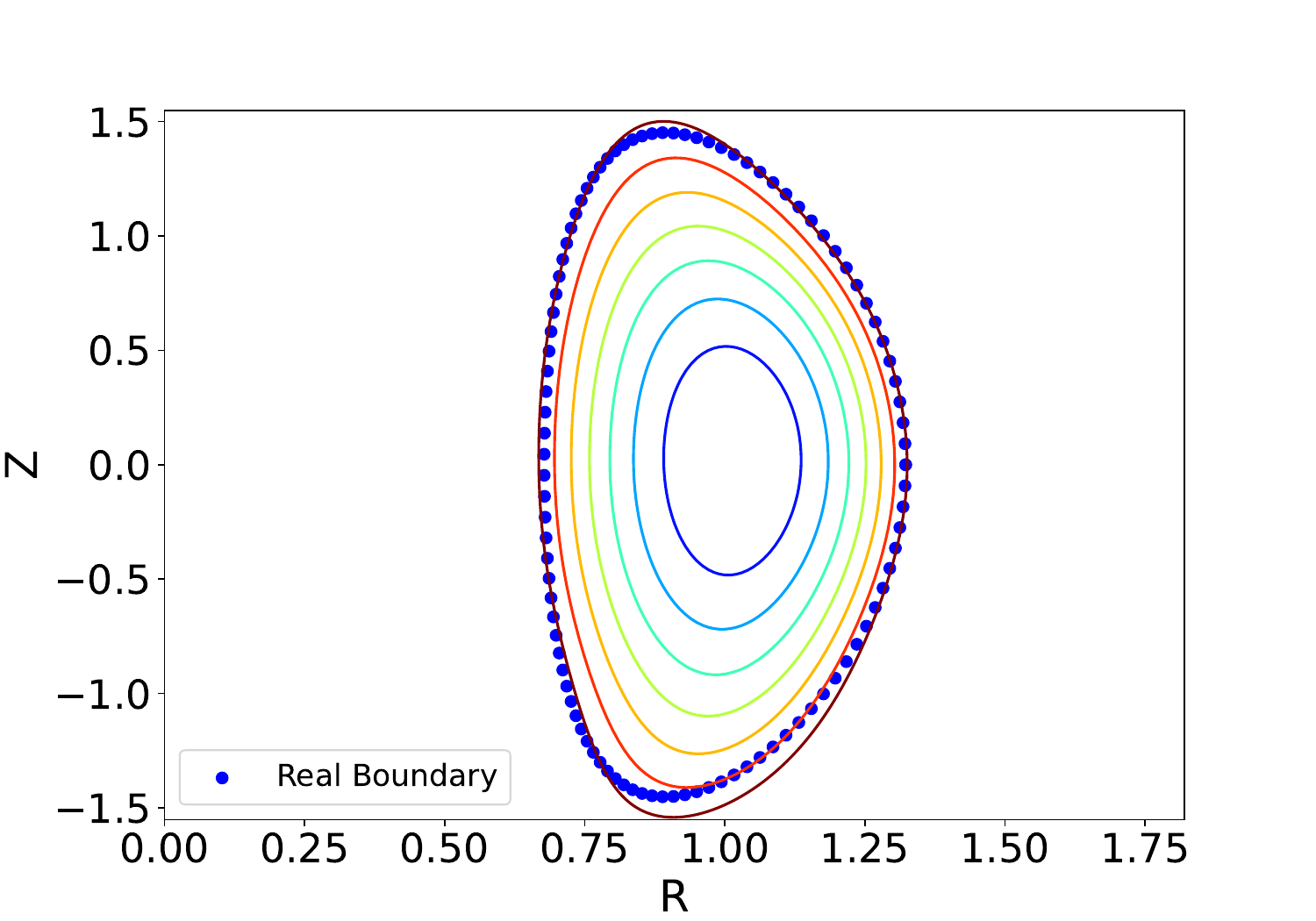}
    \caption{Predicted flux surface for $\kappa$ = 4.5 boundary vs the true boundary (blue dots).}
    \label{fig:outside_marabou}
\end{figure}

Furthermore, we extended the testing to cases beyond our trained regime, focusing specifically on the $\kappa$ range from 4.00 to 4.50. As we can see in Table \ref{tab:outside_zoneres}, Marabou was able to find a more extreme values outside our expected magnetic axis regions, indicating that there is a positional shift that is not consistent with the ground truth.  Additionally, this result aligns with how the flux surfaces visually appear, as shown in Figure \ref{fig:outside_marabou}, where the predicted boundary does not align well with the real boundary (blue dots), suggesting that the model struggles to maintain accuracy when operating outside the trained regime.

Thus, this demonstrates Marabou's capability to detect inconsistencies between the predicted and actual outputs that aligns with the actual performance of the model. This is especially beneficial when the model operates in unfamiliar or extrapolated conditions. This makes Marabou a feasible tool for detecting potential inaccuracies or shifts in predictions, reinforcing its value as a verification tool in ensuring that neural models adhere to their expected physical properties across different conditions.

\section{Conclusion}
In this study we have demonstrated further evaluation of neural models for the plasma equilibrium problem. First, we evaluated the generalization capabilities of both PINN and FNO models by testing how a singular trained model performs on an arbitrary boundaries, which includes unseen data.
While both models struggle to generalize to data outside the training region, they demonstrate strong performance within the trained region. 
Their strong performance within the trained region suggests it may not be necessary to train the model on every specific boundary to predict the equilibrium for a given shape. By spanning the training dataset over certain ranges, the network is able to produce accurate predictions even on unseen data. Furthermore, while both models achieve similar levels of accuracy, the PINN demonstrates a significant advantage in inference speed. The simpler architecture of the PINN not only enables faster predictions but also makes it more efficient to train compared to the more complex FNO architecture, which requires longer training times.

We assessed the feasibility of verifying physical properties of trained PINNs using the model checking tool Marabou. The limitation of 32-bit precision introduced marginal but manageable numerical discrepancies, which could affect properties requiring highly precise values. Nonetheless, we have demonstrated that it is feasible to verify the consistency of the magnetic axis over a sampled range of elongation of the boundary. Marabou successfully verified the consistency of the coordinates within the trained regions and identified inconsistencies when tested on untrained regions. This process is bolstered by Marabou's efficiency in detecting unsatisfiable results, establishing it as a powerful tool for proving properties through counterexamples. Moreover, we observed that splitting large queries into smaller ones significantly improves verification speed. 

With these findings we contribute to the knowledge of applying deep learning for physics, particularly in the context of plasma equilibrium modeling. Although the models were not tested in real-world scenarios with experimental data from physical instruments, our tests showcased their potential and advanced the possibilities of applying such techniques in real-world conditions. The evaluation of Marabou to verify physical properties, particularly for PINNs, brings novel insights into the potential for employing other formal verification tools in physics applications. 

\bibliographystyle{named}
\bibliography{ijcai25}

\end{document}